\documentstyle[12pt,draft]{article}

 \newcommand \be {\begin{equation}}
\newcommand \bea {\begin{eqnarray} \nonumber }
\newcommand \ee {\end{equation}}
\newcommand \eea {\end{eqnarray}}
 
 \newcommand \s {\sigma}

 \newcommand \al {\alpha}

\begin{document}

\title{Static chaos in spin glasses against quenched disorder perturbations}
\author{ Vicente Azcoiti$^{(1)}$, Eduardo Follana$^{(1)}$,
Felix Ritort$^{(2)}$.\\
{\small (1): Departamento de Fisica Teorica,}\\
{\small Facultad de Ciencias,}\\
{\small Universidad de Zaragoza,}\\
{\small 50009 Zaragoza (Spain)}\\
{\small \tt follana@cc.unizar.es, vicente@cc.unizar.es}\\
{\small (2): Departamento de Matematica Aplicada,}\\
{\small Universidad Carlos III, Butarque 15}\\
{\small Leganes 28911, Madrid (Spain)}\\
{\small \tt ritort@dulcinea.uc3m.es}}
\maketitle

\begin{center}
\vskip 10mm
\quad     Short title: Static chaos in spin glasses
\end{center}
\begin{abstract}
We study the chaotic nature of spin glasses against perturbations of the
realization of the quenched disorder. This type of perturbation modifies
the energy landscape of the system without adding extensive energy. We
exactly solve the mean-field case, which displays a very similar chaos
to that observed under magnetic field perturbations, and discuss the
possible extension of these results to the case of short-ranged
models. It appears that dimension four plays the role of a specific
critical dimension where mean-field theory is valid. We present
numerical simulation results which support our main conclusions.
\end{abstract}
\vfill
\begin{flushright}
{\bf \hfill MA/UC3M/03/95}\\
{\bf \hfill DFTUZ/95/04}\\
{\bf \hfill cond-mat/9502045}
\end{flushright}

\vfill
\newpage

\section {Introduction}

A long debated problem in spin glass theory concerns the correct
description of the statics of the low temperature phase\cite{LIBRO}.
There is wide consensus on the fact that the mean-field theory is well
understood in its essentials, while the nature of the equilibrium states
for short-ranged models is still a controversial subject. Two
compeeting pictures or approaches have been proposed: the
mean-field picture and the droplet model. The mean-field theory has
revealed enormously complex as a comprehensive approach to understand
short range models. Consequently, the search for
different approaches like droplet models \cite{DROPLETS} has been encouraged.
These models,
being phenomenological, try to capture the main aspects underlying the
equilibrium and non-equilibrium properties of short-ranged
models. Unfortunately, the mean-field way and these phenomenological
approaches are far from being complementary and much effort has been
devoted during the past years to discern what is the correct
picture. Numerical simulations have played a prominent role in this task
even though the main question still remains unsolved. The main
problem relies on the large amount of computer time needed in order
to reach the equilibrium.


Despite of the fact that both pictures are in fact contradictory
in their essentials,
there are however some common predictions in both approaches.
Since it is very difficult to decide what
is the correct picture, the strategy of searching for common features
in both pictures can be useful to shed light on this
controversy. Static chaos appears as a good starting point for this
program. By static chaos we understand the sensitivity of the low
temperature phase of spin glasses against static perturbations, like
changes in the temperature or changes in the magnetic field. Mean-field
theory \cite{KONDOR} and droplet models \cite{DROPLETS} predict that
spin-glasses, in the most general case, are chaotic. In mean-field
theory, the mechanism of chaos is due to the small free energy
differences between the different equilibrium states. These are of
order $O(1/N)$ and a small perturbation completely reshuffles the
Boltzmann weights $w_{\alpha}\sim exp(-N\beta f_{\al})$ of the
different equilibrium states ($\al$ and $f_{\al}$ stand for
equilibrium state and its free energy respectively). In droplet
models, the application of a perturbation causes a reorganization of
the spin-spin correlations at long distances. In both pictures the
system is much sensitive to the applied perturbations.

A nice example of chaos concerns the sensitivity of spin-glasses
against magnetic field perturbations \cite{KONDOR,MIO}. The chaos
exponent (to be defined in the next section) for this type of
perturbation has been computed in mean-field theory \cite{KONDOR} and
numerically measured in short-ranged models \cite{MIO}. Surprisingly,
this chaos exponent does not depend on the dimensionality of the
system \cite{MIO}. Even though we do not know a theoretical derivation
of this result, it appears to be enough sound in order to be
considered.  Droplet models can give an explanation for this
result under the assumption that 3 is the lower critical dimension in
Ising spin glasses (also a long debated problem \cite{MPR}).  In the
context of droplet models, the chaos exponent for magnetic field
perturbations is related to the thermal exponent $\theta$ which measures
the free energy cost of the droplet excitations. The result
$\theta=\frac{d-3}{2}$ implies that the chaos exponent is $2/3$ and
does not depend on the dimension. This has to be compared to the known
results, $\theta=-1$(exact) in $d=1$ \cite{HEIDELBERG}, $\theta\simeq
-0.48$ in $d=2$\cite{KAWA} and the exact result for the chaos exponent
($2/3$) in the Gaussian approximation to mean-field theory
\cite{KONDOR}
Apparently the simple
expression previously reported for $\theta$ correctly matches the
small $d$ regime to the infinite dimension result.

Regarding other type of perturbations, the situation is less clear. For
instance in the case of temperature changes, it remains unclear how much
chaotic is the system (see \cite{FRANZ} for recent results). If chaos
exists then it is certainly small and the possibility that chaos is
marginal \cite{KONDOR} cannot be excluded. Numerical results in the case
of short-range models \cite{UNP} show that chaos in temperature is also
very small, as in the mean-field case.

The perturbations previously commented share the common property that
they add energy to the system. This work is devoted to the study of a
perturbation which does not add extensive energy to the system. In
particular we will study chaoticity against changes of the realization
of the quenched disorder. Because of the self-averaging property we
expect that changes in the realization of the disorder (keeping the form
of the disorder distribution) shoud not add extensive energy to the
system. Therefore chaoticity appears because of a complete reshuffling
of the free energies of the configurations. We will show that the system
displays chaos very similarly as for the case of magnetic field
perturbations. Criticality of chaos against perturbations of the
quenched disorder has been studied by other groups \cite{BANAVAR}. The
perturbation we are interested in differs from others by the fact that
we change the sample realization without moving the system to a new
point in the phase diagram.

The paper is organized as follows.
Section 2 is devoted to the study of chaos in mean-field
models. In section 3 we discuss on the results for the short-ranged
models. Section 4 presents the scaling approach we have used to obtain
the chaos exponents and shows the numerical results. Finally we
present our conclusions in section 5.

\section{Chaos in mean-field theory}

We consider the models described by the Hamiltonian

\be
H[\s]=-\sum_{(i,j)}J_{ij}\,\s_i\s_j -h\sum_i\s_i
\label{eq1}
\ee

The couplings in (1) are Gaussian symetrically distributed random
variables with zero mean and $1/N$ variance, where $N$ is the number of
spins.  The perturbation we consider consists in changing randomly the
sign of a fraction $r$ of the couplings, i.e. for each coupling we
change its sign with probability $r$.  On average, a total number $Nr$
of the couplings $J_{ij}$ are changed to $-J_{ij}$ . In this way we
keep the new configuration of $J's$ in the same ensemble of disorder
realizations without moving the system in the phase diagram. This is
different from other type of perturbations in which, for instance (see
\cite{BANAVAR}), the $J_{ij}$ are changed by a small amount
$\delta\cdot x_{ij}$ where $x_{ij}$ is a random number and $\delta$ is
small. In this case the variance of the distribution is increased (it
grows proportionally to $\delta^2$) and we add energy to the system. In
what follows we will consider, for simplicity, the case of zero magnetic
field.

Denoting by $R$ the set of couplings which change sign, then we can
write the perturbed Hamiltonian as,

\be
H_r[\s]=
-\sum_{(i,j)}\,J_{ij}\s_i\s_j\,+2\sum_{(i,j)\in R}\,J_{ij}\s_i\s_j
\label{eq2}
\ee

The sum runs over nearest neighbors in a lattice of dimension $d$. The
mean-field case can be obtained in several ways. In the infinite-range
model or SK model, all the spins interact one to the
other. Alternatively, one can consider the finite connectivity random
lattices (with fixed or average number of neighbors \cite{DOMINICIS}).

Once we have defined the perturbation we construct a full Hamiltonian
$H_{12}[\s,\tau]$ defined in a space of two sets of variables
$\lbrace\s_i,\tau_i;i=1,..N\rbrace$. The Hamiltonian $H_{12}$ is the sum
of the unperturbed Hamiltonian $H[\s]$ plus the perturbed Hamiltonian
$H_r[\tau]$,

\be
H_{12}[\s,\tau]=H[\s]+H_r[\tau]
\label{eq3}
\ee

We define the usual spin-glass correlation functions,

\be
G(x)=\overline{\langle\s_0\tau_0\s_x\tau_x\rangle}
\label{eq4}
\ee

where $\overline{(..)}$ means average over the quenched disorder and
$\langle..\rangle$ corresponds to the thermal average over the full
Hamiltonian $H_{12}$. The degree of coherence of the two systems is
measured by the overlap function,

\be
P(q)=\overline{\langle\delta(q-\frac{1}{N}\sum_i\s_i\tau_i)\rangle}
\label{eq4b}
\ee

At large distances $G(x)$ behaves like

\be
G(x)\sim exp(-x/\xi(r))~~~~~~~~
\label{eq5}
\ee

where $\xi(r)$ is the chaos correlation length, which is finite for a finite
perturbation $r$ (we identify the perturbation with the fraction $r$
of changed couplings). The chaos correlation length diverges when $r\to
0$ if the unperturbed system stays in the spin-glass phase, including the
critical point. This is because in the limit $r\to 0$, $G(x)$
converges to the usual spin-glass correlation function which has
infrared singularities due to the existence of zero modes. The chaos
correlation length diverges like,

\be
\xi(r)\sim r^{-\lambda}
\label{eq6}
\ee

where $\lambda$ is the {\em chaos exponent} which can be exactly
computed in some particular cases. From this definition it is clear that
static chaos is absent when the system stays at a temperature above the
spin-glass transition (the paramagnetic phase). But this assertion is
true only if $\xi(r)$ smoothly converges to the finite correlation
length at that temperature when $r\to 0$. This is the case in mean-field
theory but should not be necessarily true in finite dimensions
\cite{NIFLE}. The exponent $\lambda$ can also depend in principle on the
temperature. We will show in mean-field theory that $\lambda$ is
constant in the low $T$ phase. Furthermore, at the critical point, we
expect $\lambda$ to depend on the critical exponents, even though this
is not always the case\footnote{For the problem of chaos in temperature
there is a new chaos exponent independent of the usual critical
exponents \cite{HILHORST}.}.

Now we face the problem of computing the exponent $\lambda$ in
mean-field theory. We follow the standard procedures (see \cite{MIO} for
details) and we apply the replica method to the full Hamiltonian
$H_{12}$ eq.(\ref{eq3}),

\be
\beta f=\lim_{n\to 0} \frac{log(\overline{Z_J^n})}{nN}
\label{eq7}
\ee

Introducing Lagrange multipliers for the different order parameters one
gets a saddle point integral

\be
\overline{Z_J^n}=\int\,dP\,dQ\,dR\, exp(-N A[PQR])
\label{eq8}
\ee

with

\be
A[PQR]=\frac{\beta^2}{2}\sum_{a<b}(P_{ab}^2+Q_{ab}^2)+
\frac{\beta^2}{2}\sum_{a,b}(R_{ab}^2)-log(Tr_{\s\tau} exp(L))
\label{eq9}
\ee

where $a,b$ denote replica indices which run from 1 to $n$, and

\be
L[\s,\tau]=\beta^2\sum_{a<b}(Q_{ab}\s_a\s_b+P_{ab}\tau_a\tau_b)+
\beta^2\sqrt{1-2r}\ \sum_{a,b}(R_{ab}\s_a\tau_b)
\label{eq9b}
\ee
There is one stable solution to the equations of motion,

\be
Q_{ab}=P_{ab}=Q_{ab}^{SK}~~~~~~~~~~~~R_{ab}=0
\label{eq10}
\ee

where $Q_{ab}^{SK}$ is the solution for the unperturbed system.  The
 order parameter $R_{ab}$ measures the degree of correlation
 eq.(\ref{eq4b}) between the two systems via the relation,

\be
P(q)=\frac{1}{n^2}\sum_{a,b}\,\delta(q-R_{ab})
\label{eq10b}
\ee

The stability of the solution $R=0$ means that there is chaos
against coupling perturbations. This is indeed very similar to the case
of chaos in a magnetic field. Now we can compute, in the Gaussian
approximation, the correlation function $G(x)$ of eq.(\ref{eq5}). The
computations can be easily done in Fourier space. We define,

\be
C(p)= \sum_x G(x) e^{ipx}
\label{eq11}
\ee

In order to find $C(p)$ we need to compute the spectrum of fluctuations
in the direction $R_{ab}$ around the stable solution (eq.(\ref{eq10})).
The full expression has been
reported in \cite{KONDOR}. Its singular part is given by
\be
C(p)=\int_{0}^{q_{max}}\,dq\,\int_{0}^{q_{max}}\,dQ
\frac{p^2+1+\alpha(q)\alpha(Q)}{(p^2+1-(1-2r)\alpha(q)\alpha(Q))^3}\\
\label{eq12}
\ee
with
\be
\alpha(q)=\beta(1-q_{max}+\int_{q}^{q_{max}}\,dq\,x(q))
\label{eq13}
\ee

where $\beta$ is the inverse temperature and $q(x)$ is the order
parameter function associated to the spin-glass. This expression yields
the singular behavior of the correlation function in the spin-glass
phase \cite{MIO},

\be
C(p)\sim p^{-4},~~~~~p\to 0
\label{eq13b}
\ee

The chaos correlation length $\xi(r)$ is given by the minimum
eigenvalue of the stability matrix,

\be
\lambda_{min}=2\beta^2 r
\label{eq14a}
\ee

This yields,

\be
\xi(r)\sim\lambda_{min}^{-\frac{1}{2}}\sim r^{-\frac{1}{2}}
\label{eq14}
\ee

This result is valid at and below the critical point.  We expect it to
be valid also in other mean-field models like, for instance, finite
connectivity random lattices.  In this case, where analytical
calculations become much more involved, we expect to obatin the same
results. This will be nicely corroborated by our numerical simulations
in section 4.

\section{Chaos in short-ranged systems}

Now we face the problem of extending our results to finite
dimensions. This is a non trivial task and we will present a derivation
only for the one-dimensional case.

The chaos exponent can be exactly computed in one dimension since we
know how to construct the ground state in this case. The exponent
$\lambda$ is a zero-temperature exponent because there is no
phase transition at finite $T$.
The Hamiltonian reads,

\be
H=-\sum_i\,J_i\s_i\s_{i+1}
\label{eq15}
\ee

The perturbation consists in changing the sign of a fraction $r$ of
the couplings in a random way, and we assume that the $J's$ are
distributed around $J=0$ with a finite weight at $J=0$ (this is
essential for the scaling arguments). When a fraction of the couplings
is changed, the new ground state is constructed inverting domains close
to the deffects. The energy excess of these deffects scales like $rL$,
where $L$ is the length of the spins chain.  On the other hand,
domain excitations (in this case these excitations are inversions of
compact domains) scale like $L^{\theta}$ with $\theta=-1$, where
$\theta$ is the thermal exponent introduced in droplet models (see
\cite{HEIDELBERG} for a derivation.).  This gives the chaos correlation
length,

\be
\xi\sim r^{-\frac{1}{2}}
\label{eq16}
\ee

Domains of lenth $L$ above this characteristic length $\xi$ are
destroyed by the bond deffects. Below the characteristic length, the
domains are nearly insensitive to the perturbation. This result is valid
for a distribution of couplings with finite weight at $J=0$
\footnote{For a distribution of couplings with zero weight at $J=0$ one
finds $\xi\sim r^{-1}$}.  We have found therefore for the chaos
correlation length the same result as in mean-field theory
. Unfortunately we cannot do more in order to compute this exponent in
other dimensions because we do not know the structure of the ground
state.  Anyway, we can try to estimate the energy the excitations after
applying the perturbation.
In the more general case, a perturbation of the type we are studying
here will modify the ground state energy by a quantity proportional to
the number $r$ of created defects (because the total fraction of
frustrated and unfrustrated bonds is finite). In addition, we can
suppose that this energy will scale like the size of the system
$L^{\alpha}$ with $\alpha\le d$. In principle, the exponent $\alpha$ is
unknown and we do not know how to estimate it. If we assume
$\alpha=\frac{d+1}{2}$ (and using $\theta=\frac{d-3}{2}$), this gives
the dimension independent result $\xi\sim r^{-\frac{1}{2}}$.
Unfortunately we are unable to estimate the exponent $\alpha$ and a
numerical computation of this exponent in 2 dimensions would be very
interesting.

\section{Numerical results}

In this section we will discuss on our Monte Carlo simulations in order
to test the results obtained in the previous sections for the chaos
exponents. Furthermore, we will present simulations in four
dimensions. Our results are compatible with the fact that the chaos
exponents in finite dimensions are compatible with the mean-field ones.

We have simulated two types of mean-field models (the Sherrington
Kirkpatrick -SK- model \cite{SK} and the random finite connectivity
lattice model \cite{DOMINICIS}) and a four-dimensional (4d) Ising spin
glass for which the existence of a finite $T$
phase transition is well established  \cite{BADONI}. Monte Carlo
simulations implement the
Metropolis algorithm (for the mean-field models) and the heat-bath
algorithm (in the 4d case). Special attention has been payed in order
to thermalize the samples.

\subsection{The finite-size scaling approach}

In order to measure the chaos exponents, we have performed a finite-size
scaling analysis \cite{MIO}. The idea is to compute the overlap between
two copies of the system, one copy with an initial realization of the
disorder, the other one with the perturbed realization. The overlap is
defined as,

\be
q=\sum_{i=1}^N\,\s_i\tau_i~~~~~~~
\label{eq17}
\ee

\noindent We define the {\em chaos parameter} a($r$),

\be
a(r)=\frac{\overline{\langle\s_i\tau_i\rangle^2}_r}
{\overline{\langle\s_i\tau_i\rangle^2}_{r=0}}
\label{eq18}
\ee

i.e. we normalize the correlation between the unperturbed and the
perturbed system to the autocorrelation of the unperturbed system. In
this way $a(0)=1$ by definition. The system is chaotic if the quantity
$a(r)$ (in the thermodynamic limit) jumps to 0 as soon
as $r$ is finite. This means that,

\be
\lim_{r\to 0}\lim_{N\to\infty} a(r)=0~~~~~~~while~~~~~a(0)=1
\label{eq19}
\ee

It is crucial to perform the limits in the order previously
indicated. Since $a$ is an adimensional quantity, we expect it will
scale like

\be
a\equiv f(L/\xi)
\label{eq20}
\ee

where $\xi$ is the chaos correlation length of eq.(\ref{eq5}). In the
mean-field case we find for the spin-glass phase (using
eqs. (\ref{eq13b}) and (\ref{eq14}))

\be
a\equiv f(Nr^2)~~~~~~.
\label{eq21}
\ee

and at the critical point we get (using the singular behavior
$C(p)\sim p^{-2}$ for $p\to 0$)

\be
a\equiv f(Nr^3)~~~~~~.
\label{eq22}
\ee

In the case of short-range models we can derive the scaling behavior using
equation (\ref{eq16})

\be
a\equiv f(rL^2)
\label{eq23}
\ee

Since only one exponent (the chaos exponent) must be fitted,
these scaling relations are highly predictive.
Comparing equations (\ref{eq21})
and (\ref{eq23}) we observe that $d_u=4$ plays the role of a specific
critical dimension. The situation is the same as in the case of magnetic
field perturbations \cite{MIO}, where the value of this dimension only
depends on the behavior of the propagator $C(p)$ in the limit $p\to 0$.
In this case we expect the scaling functions $f(x)$ in equations
(\ref{eq21}) and (\ref{eq23}) to coincide except by the presence of some
logarithmic corrections. A numerical test of this prediction is shown in the
following subsections.

\subsection{Numerical results in mean-field models}

The SK model is defined by the following hamiltonian

\be
H=\sum_{i<j}\,J_{ij}\s_i\s_j
\label{eq24}
\ee

with the $J_{ij}$ distributed according to the function $p(J_{ij})$. In
the thermodynamic limit, the only relevant feature of the $p(J_{ij})$ is
its variance (we restrict to distributions with zero mean). To speed up
the numerical computations we have taken a binary distribution of
couplings, i.e. the $J's$ can take the values $\pm \frac{1}{\sqrt{N}}$
with equal probability.

We have simulated the SK model at the critical temperature $T=1$ and
below the critical temperature. We have computed the chaos parameter $a$
for different values of $r$ (tipically $r$ runs from 0 to
0.5). Simulations were done for lattice sizes ranging from $N=32$ to
$N=1000$. Figures 1 and 2 show the scaling laws eq.(\ref{eq21}) and
(\ref{eq22}) at the critical point $T=1$ and below the critical point
$T=0.7$ respectively. Data do nicely fit the predictions. We have also
simulated the random finite-connectivity lattice model(FC model).  In
this model each point of the lattice is connected (in average) to a
finite number $c$ of neighbors \footnote{One can also consider
the case in which the connectivity is fixed and equal to $c$}.  In
this case the FC model model is defined by,

\be
H=\sum_{i<j}\,J_{ij}\s_i\s_j
\label{eq25}
\ee

where the $J_{ij}$ are distributed according to,

\be
{\cal P}(J_{ij})=\frac{c}{N}\,p(J_{ij})+(1-\frac{c}{N})\delta(J_{ij})
\label{eq26}
\ee

and $p(J_{ij})$ is given by,

\be
p(J_{ij})=\frac{1}{2}\delta(J_{ij}-1)+
\frac{1}{2}\delta(J_{ij}+1)
\label{eq27}
\ee

The parameter $c$ is the average connectivity of the lattice. This model
can be exactly solved, the only difference with respect to the SK model
being that there appear an infinite set of order parameters, which can
be absorbed in a global order parameter \cite{DOMINICIS}.  The model has
a phase transition at a temperature $\beta_c$ given by,

\be
1=(c-1)\int_{-\infty}^{\infty}p(J)tanh^2(\beta_c\,J)=(c-1)tanh^2(\beta_c)
\label{eq28}
\ee
This expression implies that to have a phase transition, we need
$c>2$. To compare with
the results of the 4d case, we have simulated the FC model with $c=8$, in
order to have the same number of nearest neigbors than the 4d
model. The transition temperature is in this case $T_c\simeq 2.76$. We have
simulated this model at the critical temperature and below that
temperature, at $T=2.0$. The results for the chaos parameter $a$ are
shown in figures $3$ and $4$. The agreement with the scaling predictions
(eqs.(\ref{eq21}) and (\ref{eq22})) is also fairly good.

\subsection{Numerical results in four dimensions}

We have also done numerical simulations of the Ising spin glass model in four
dimensions with the purpose of analysing the dimensionality
effects on the chaos
exponent. We have considered the Ising spin glass at $d=4$ because it is
widely accepted that there is a finite $T$ phase transition in this case
\footnote{In the three dimensional case there is still much controversy
on the existence of a finite $T$ transition\cite{MPR}}.

We have simulated the model (eq.(\ref{eq1})) with a nearest neighbour
interaction, periodic boundary conditions and using a
discrete binary distribution of couplings as in
eq.(\ref{eq27}). We expect to
obtain the same results as in the case of a continuous distribution of
couplings with $p(J=0)$ finite. The model has a transition at $T_c\sim
2.05$ \cite{BADONI}. We have done simulations at $T=T_c$ and
$T=1.7$. The results are shown in figures 5 and 6.

At the critical point we obtain a chaos exponent $\lambda\sim
\frac{2}{3}$. It is not clear to us how to obtain this exponent in
terms of the critical exponents and if it represents a new critical
chaos exponent.

The results in figure 6 show that eq.(\ref{eq23}) is in pretty good
agreement with the data. Now we will show that four dimensions is well
compatible with the upper critical dimension for the criticality of
chaos. In order to get this result, we will compare the different values
of the chaos parameter $a$ for different sizes with the corresponding
values of the FC model with $c=8$. We compare with the FC model, instead
of the SK model, because we expect that logarithmic corrections, if
present, should be smaller in the FC model than in the SK model. Both
are mean-field models even though the FC model resembles the finite $d$
model much more than does the SK model. This fact should reflect in the
nature of the corrections to the universal mean-field behavior. It is
clear that in order to compare the FC model with the four dimensional
model we have to put the system in equivalent points within the
phase diagram. We expect the universal function $f(x)$ to depend on the
temperature (which is an external parameter) in the following way

\be
a\equiv f(A(T) (L/\xi)^d)
\label{eq29}
\ee

In four dimensions, the scaling function $f$ still depends on the
temperature via the universal amplitude $A(T)$.  It is reasonable to
assume that the dependence of the amplitude $A(T)$ on the temperature
enters through the spin-glass order parameter $q(T)$. More concretely,
below but close to $T_c$ we expect,

\be
A(T)\sim q^2(T)
\label{eq30}
\ee

because the argument of the scaling function $f$ of equation
(\ref{eq29}) scales like the singular part of the free energy which in
mean field theory scales like $Q_{ab}^2$ (see eq.(\ref{eq9})).
Consequently we have to normalize the adimensional ratio $(L/\xi)^d$ to the
corresponding value of the Edwards-Anderson order parameter for that
temperature. For $N=256$ the FC model gives $q(T=2.0)\simeq 0.11$ and
the 4d model at $L=4$ gives $q(T=1.5)\simeq 0.25$, the ratio of both numbers
being $2.5$. Simulation data for both models are shown in figure 7. If
one considers the SK model then one observes that data fits well but not
so nicely as in the case of the FC model.

\section{Conclusions}

We have investigated the sensitivity of spin glasses against the
application of a particular static perturbation. In particular, we have
studied the nature of the static chaos when a perturbation to the
realization of the quenched disorder is applied to the system.  This can
be done in several ways. In our case we have considered a perturbation
wich, in average, does not add energy to the system. Due to the
self-averaging property we expect that a change in the sign of a finite
fraction of the total number of couplings in the system should not
change its mean statistical properties (and in particular, its energy).
This makes the new perturbed system to stay in the same point in the
phase diagram. The existence of strong chaos for this type of
perturbation proves that the reshuffling of the Boltzmann weights of the
different states is complete. This differs from the case where the
perturbation consists in applying a magnetic field to the system or
where its temperature is changed. In these cases extra energy is
supplied to the system.

We have solved the mean field theory and we have extracted the chaos
exponent for this type of perturbation. The analytical solution of this
problem is very similar to that of chaos against magnetic field
perturbations where the chaos correlation length can be exactly computed
\cite{KONDOR}. This is in contrast to what happens when the temperature
is changed. In the last case the system is much robust against the
perturbation and a high degree of correlation between the configurations
at both temperatures is preserved \cite{FRANZ}.

We have observed that the mean-field chaos exponent $1/2$ in the
spin-glass phase is exact also at one dimension. A finite-size scaling
approach to the criticality of chaos shows that $d=4$ plays the role of
an upper critical dimension for the chaos problem. Finite-size scaling
studies are very powerful in order to get the chaos exponents. This is
because we only need to determine one free parameter to make the data
corresponding to different sizes to collapse in a unique scaling
function. We have performed numerical simulations of mean-field models
which are in agreement with the theory. Simulations in four dimensions
are in very good agreement with the fact that $4$ plays the role of an
upper critical dimension for the criticality of chaos (see figure
7). Furthermore, the fact that the mean-field chaos exponent is also
exact in one dimension suggests that mean-field theory is probably
correct at any dimension. This is indeed very similar to what happens in the
case of magnetic field perturbations.

Finally we would like to point out two possible extensions of this work.
Firstly it would be interesting to make dynamical studies
of the relaxation of the overlap function against this type of
perturbation (as done for the remanent magnetization after
application of a magnetic field). We expect to see aging effects as in
the case of magnetic field perturbations. Secondly, it would be
interesting to extend the study of chaos to the metastable states using
the TAP formalism. Most probably, similar chaotic properties
will be observed in the structure of the metastable states.

\section{\bf Acknowledgements}

E. F. acknowledges Gobierno de Navarra through a predoctoral grant
for financial support. We thank also the $CICYT$ institution for partial
financial support.

\vfill
\newpage
{\bf Figure Captions}
\begin{itemize}

\item[Fig.~1] Chaos in the SK model at the critical point $T_c=1$.

\item[Fig.~2] Chaos in the SK model at $T=0.7$ in the spin glass phase.

\item[Fig.~3] Chaos in the FC model with $c=8$ at the critical point
$T_c\simeq 2.76$.

\item[Fig.~4] Chaos in the FC model with $c=8$ at $T=2.0$ in the spin
glass phase.

\item[Fig.~5] Chaos in the 4d Ising spin glass at the critical temperature
$T_c\simeq 2.05$. We obtain $\lambda\sim 2/3  $ for the chaos exponent.

\item[Fig.~6] Chaos in the 4d Ising spin glass at $T=1.7$ in the
spin-glass phase. The mean-field chaos exponent $\lambda=\frac{1}{2}$
fits data very well.

\item[Fig.~7] Chaos in the 4d Ising spin glass at $T=1.7$ compared to
the FC model with $c=8$. The abcisa $x$ corresponds to $Nr^2$ (with
$N=L^4$ in four dimensions).  This scaling suggests that four dimensions
is the upper critical dimension for the criticality of chaos.

\end{itemize}

\begin{thebibliography}{99}

\bibitem{LIBRO} For general reviews on spin glasses see:
M.~M\'ezard, G.~Parisi and M.~A.~Virasoro, {\em Spin Glass Theory and
Beyond} (World Scientific, Singapore 1987); K. H. Fischer and J. A.
Hertz, {\em Spin Glasses} (Cambridge University Press 1991);
G.~Parisi, {\em Field Theory, Disorder and Simulations} (World
Scientific, Singapore 1992); K. Binder and A. P. Young, {\em Spin
Glasses: Experimental Facts, Theoretical Concepts and Open Questions}
Rev. Mod. Phys. {\bf 58}, 801 (1986).

\bibitem{DROPLETS} W. L. McMillan, {\em Scaling Theory of Ising
Spin Glasses} J. Phys. C {\bf 17} (1984) 3179; A. J. Bray and
M. A. Moore, {\em The Nature of the spin-glass phase and Finite-Size
effects} J. Phys. C {\bf 18} (1985) L699; D. S. Fisher and
D. A. Huse, {\em Equilibrium Behavior of the spin-glass ordered phase}
Phys. Rev. B {\bf 38} (1988) 386; D. S. Fisher and
D. A. Huse, {\em Non-Equilibrium dynamics of spin-glasses}
Phys. Rev. B {\bf 38} (1988) 373.

\bibitem{KONDOR} I. Kondor, {\em On Chaos in spin glasses} J. Phys. A
{\bf 22} (1989) L163.

\bibitem{MIO} F. Ritort, {\em Static Chaos and Scaling Behavior in the
Spin-Glass Phase} Phys. Rev. B. {\bf 50} (1994) 6844.

\bibitem{MPR} E. Marinari, G. Parisi and F. Ritort, {\em On the 3d
Ising spin glass} J. Phys. A {\bf 27} (1994) 2687 and refernces therein

\bibitem{HEIDELBERG} A. J. Bray and M. A. Moore in
{\sl Heidelberg Colloquium in Spin Glasses}, Springer Lecture
Notes in Physics, Vol. 275 (1986).

\bibitem{KAWA} N. Kawashima, N. Hatano and M. Suzuki, {\em Critical
Behavior of the two-dimensional EA model with a Gaussian bond distribution}
J. Phys. A {\bf 25} (1992) 4985.

\bibitem{FRANZ} S. Franz and M. N. Nifle, {\em On Chaos in mean-field
spin glasses} cond-mat {\bf 9412083}.

\bibitem{UNP} F. Ritort,  unpublished results.

\bibitem{BANAVAR} M. Cieplak and J. R. Banavar, {\em Scaling and Phase
Transitions in Random Systems} in Statistical Physics (StatPhys 18) (1992)
North Holland.

\bibitem{DOMINICIS} Y. Y. Goldschmidt and C. De Dominicis, {\em
Replica Symmetry Breaking in the spin-glass model on lattices with finite
connectivity} Phys. Rev. B {\bf 41} (1990) 2184 and references therein.

\bibitem{NIFLE} M. Nifle and H. J. Hilhorst, {\em } J. Phys. A {\bf 24}
(1991) 2397.

\bibitem{HILHORST} M. Nifle and H. J. Hilhorst, {\em New Critical-Point
Exponent and New Scaling Laws for Short-Ranged Ising Spin Glasses}
Phys. Rev. Lett. {\bf 68} (1992) 2992.

\bibitem{SK} D. Sherrington and S. Kirkpatrick, {\em Infinite Ranged
Models of Spin Glasses} Phys. Rev. B {\bf 17} (1978) 4384.

\bibitem{BADONI} D. Badoni, J. C. Ciria, G. Parisi, J. Pech, F. Ritort
and J. J. Ruiz, {\em Numerical Evidence of a Critical Line in the 4d Ising
Spin Glass} Europhys. Lett. {\bf 21} (1993) 495.


\end{thebibliography}
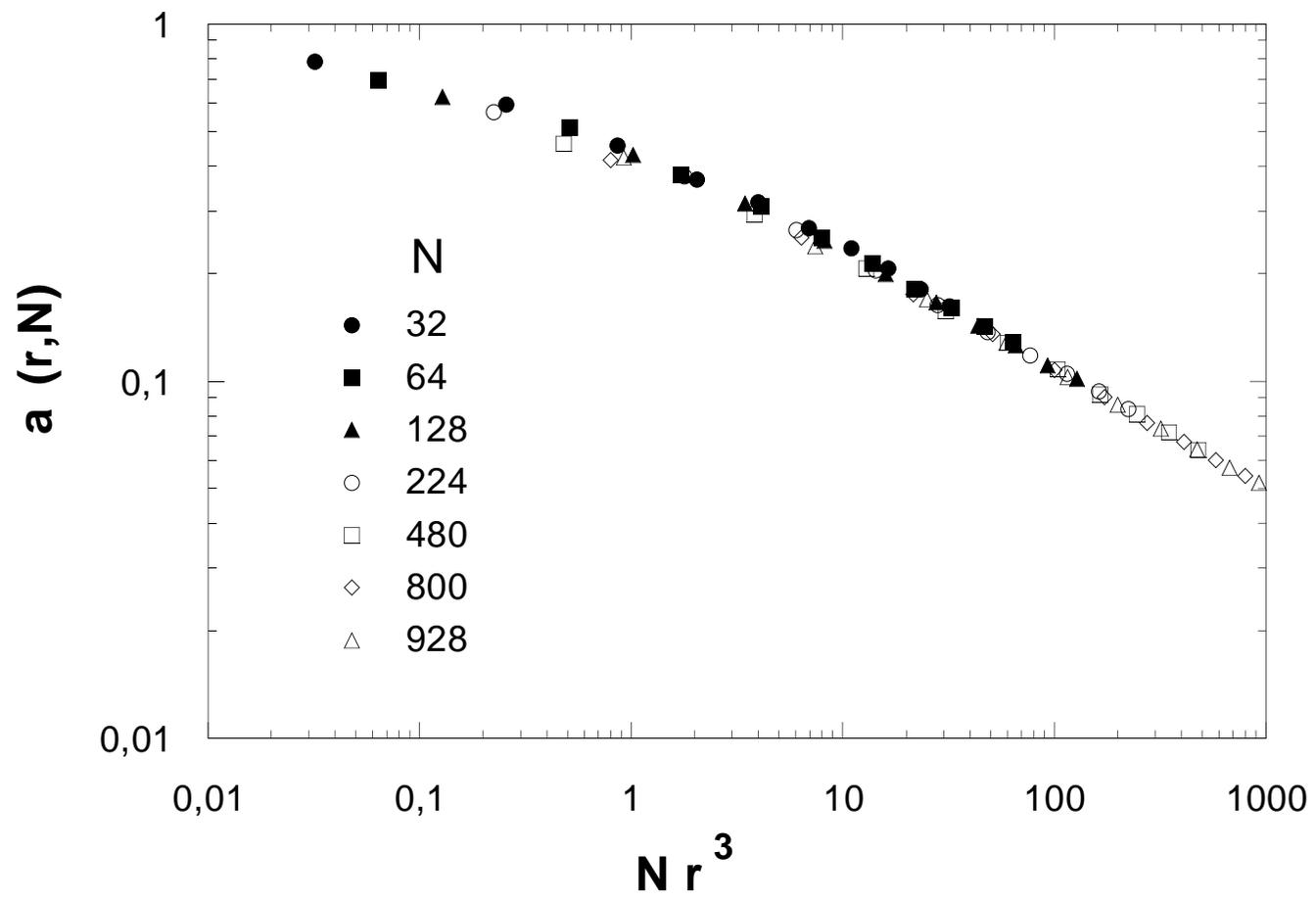
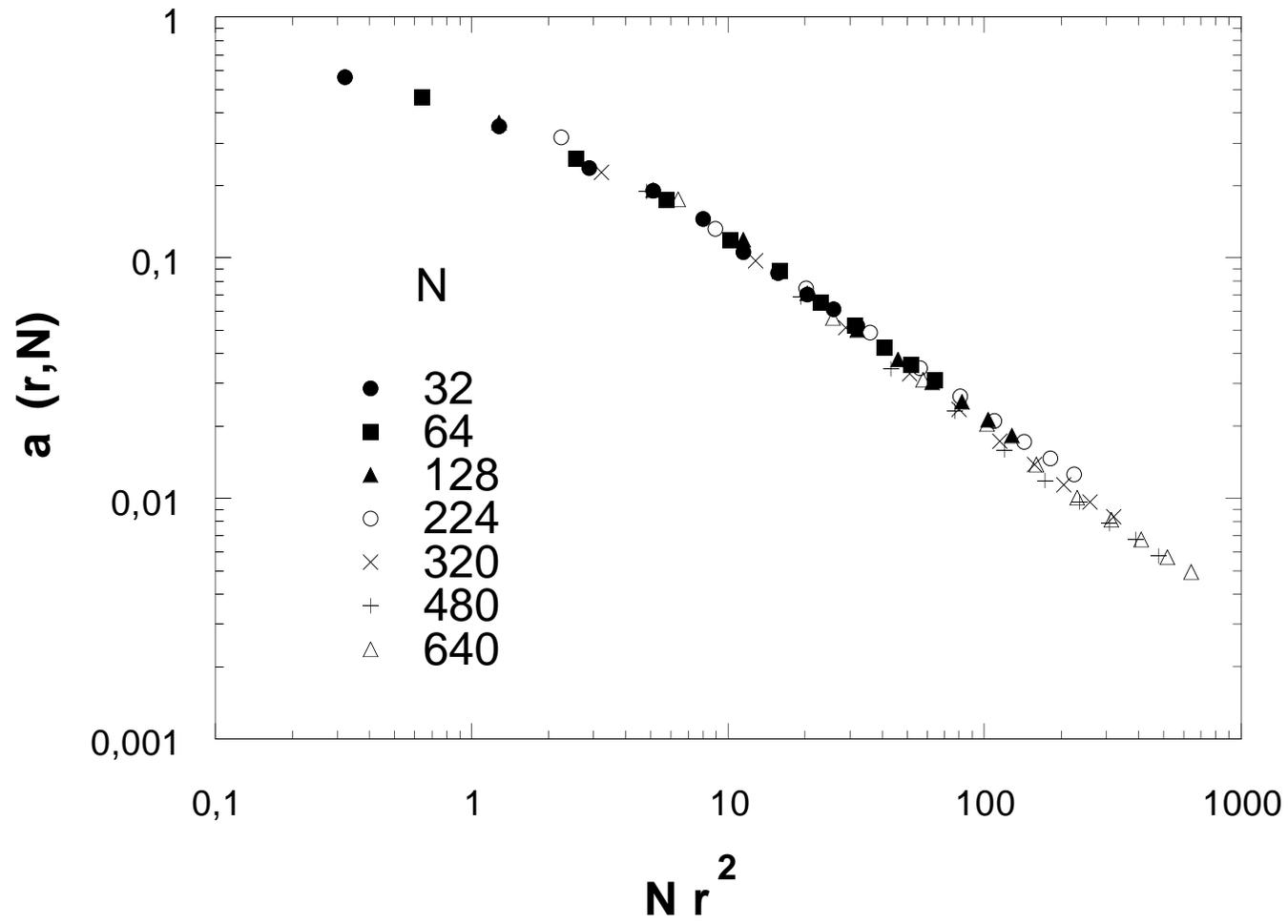
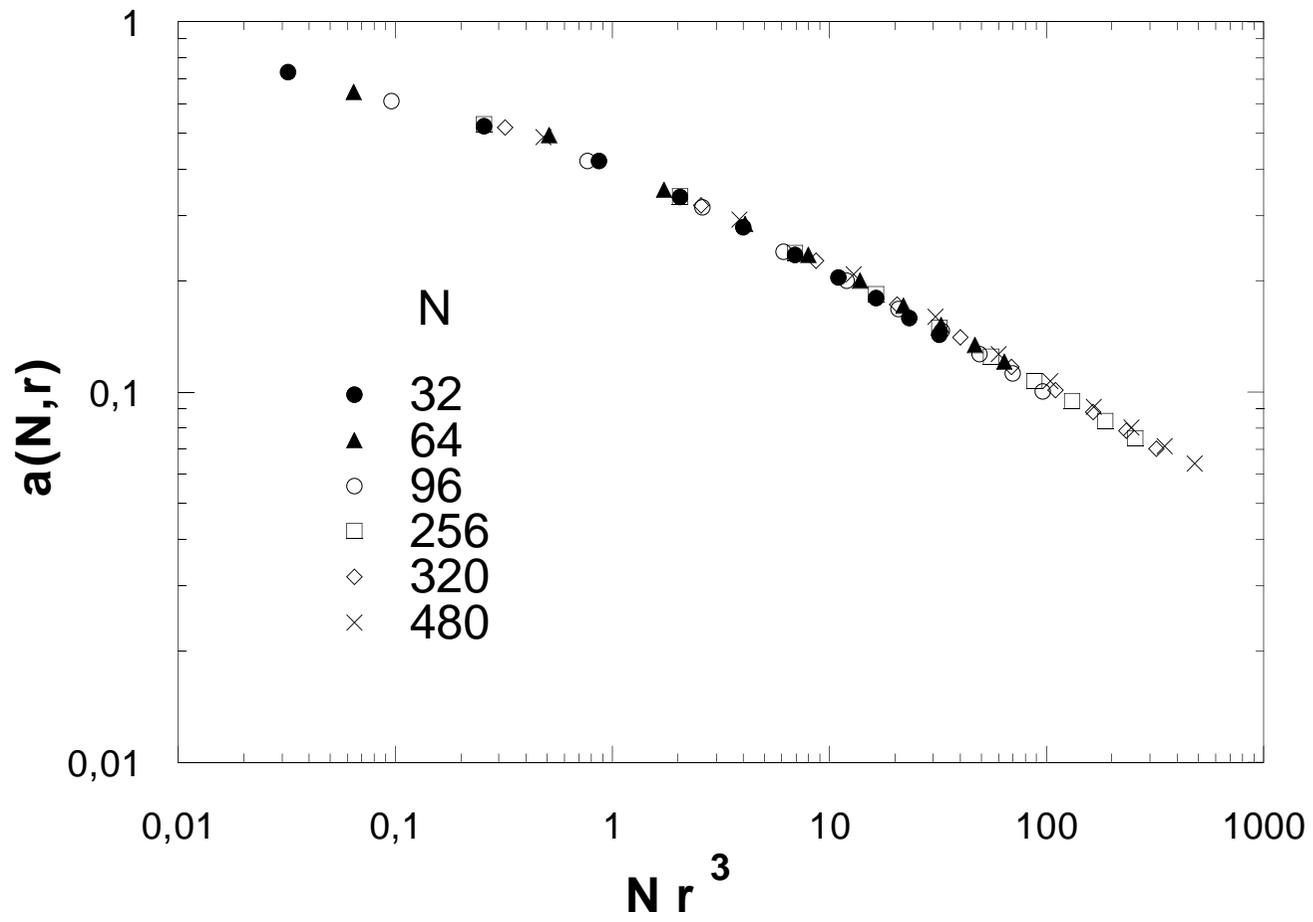
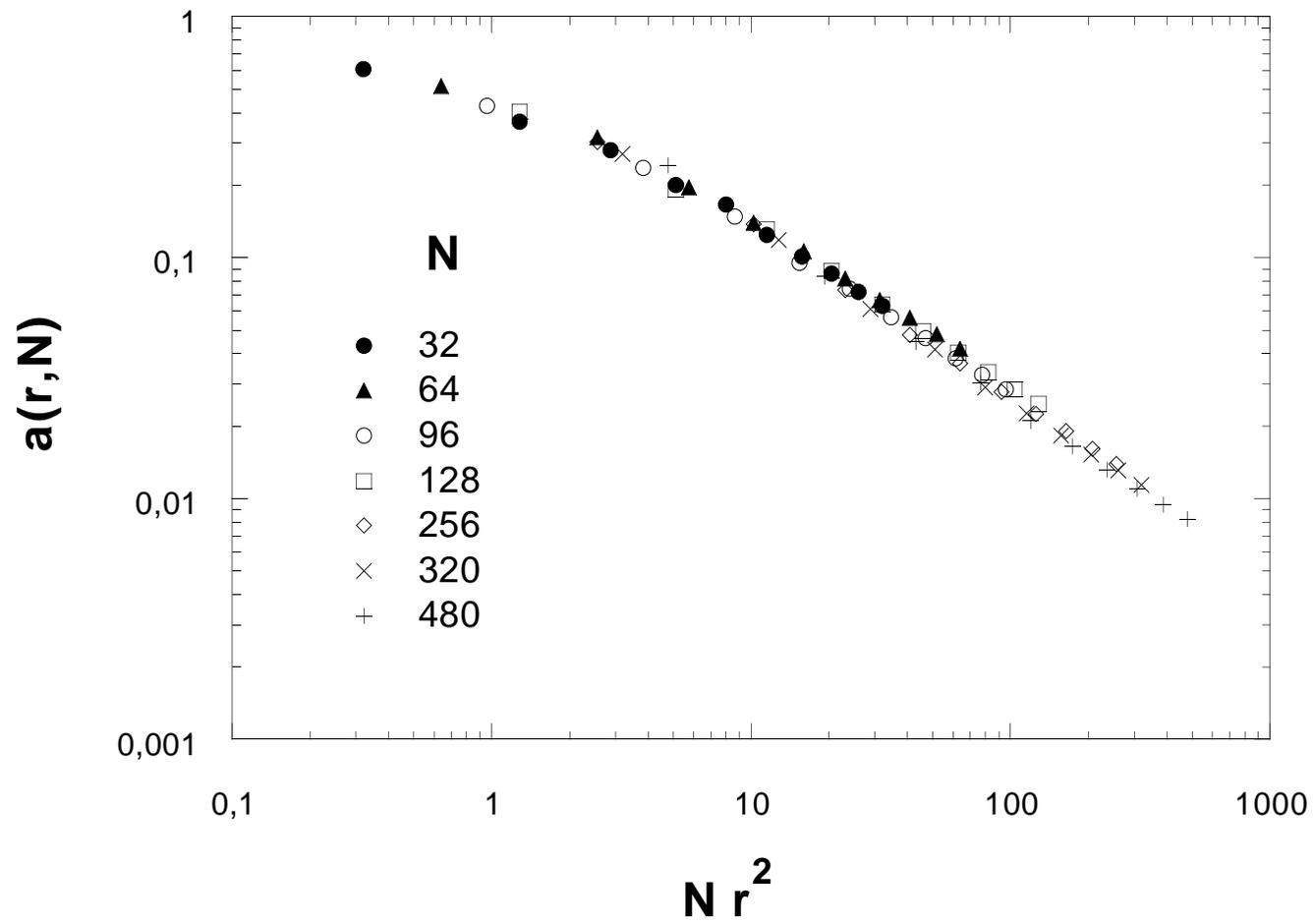
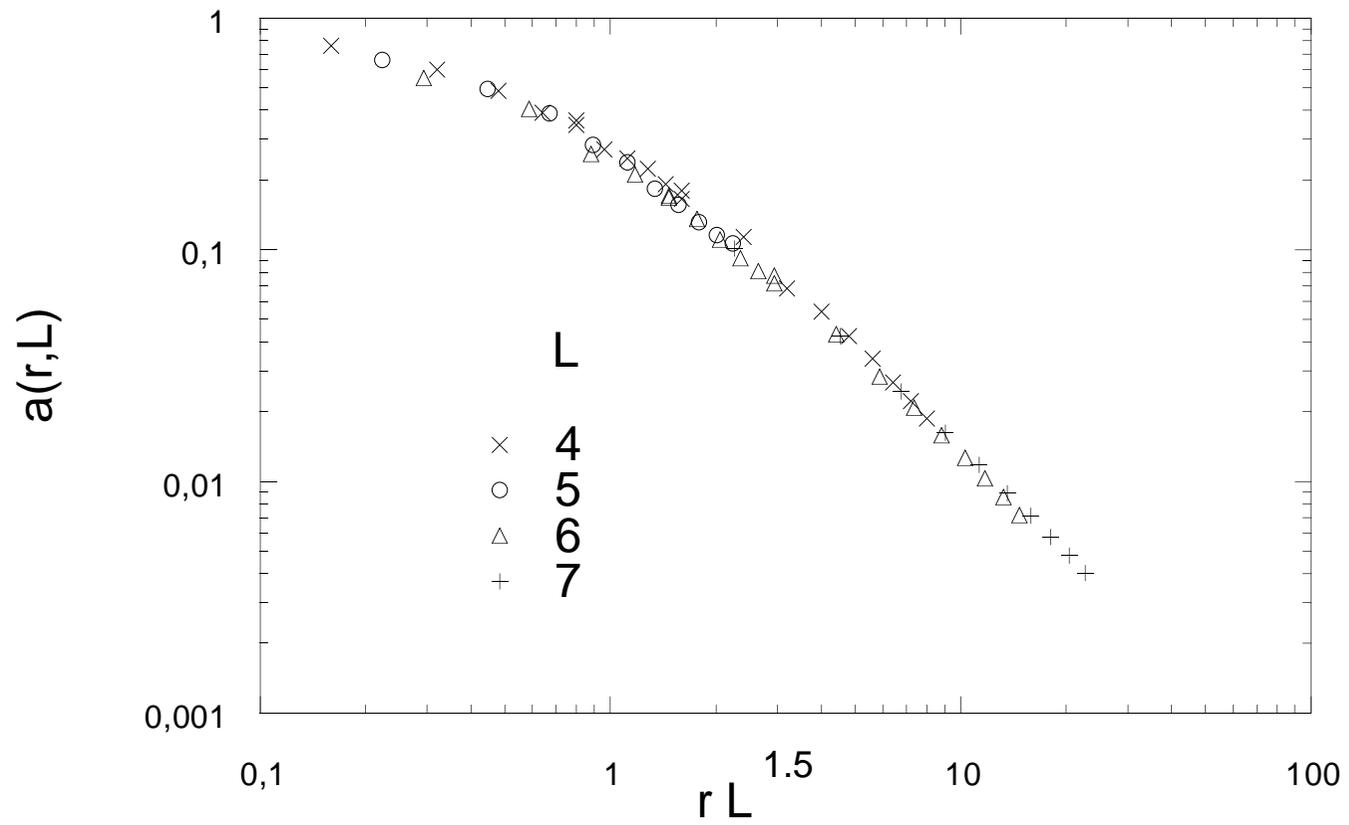
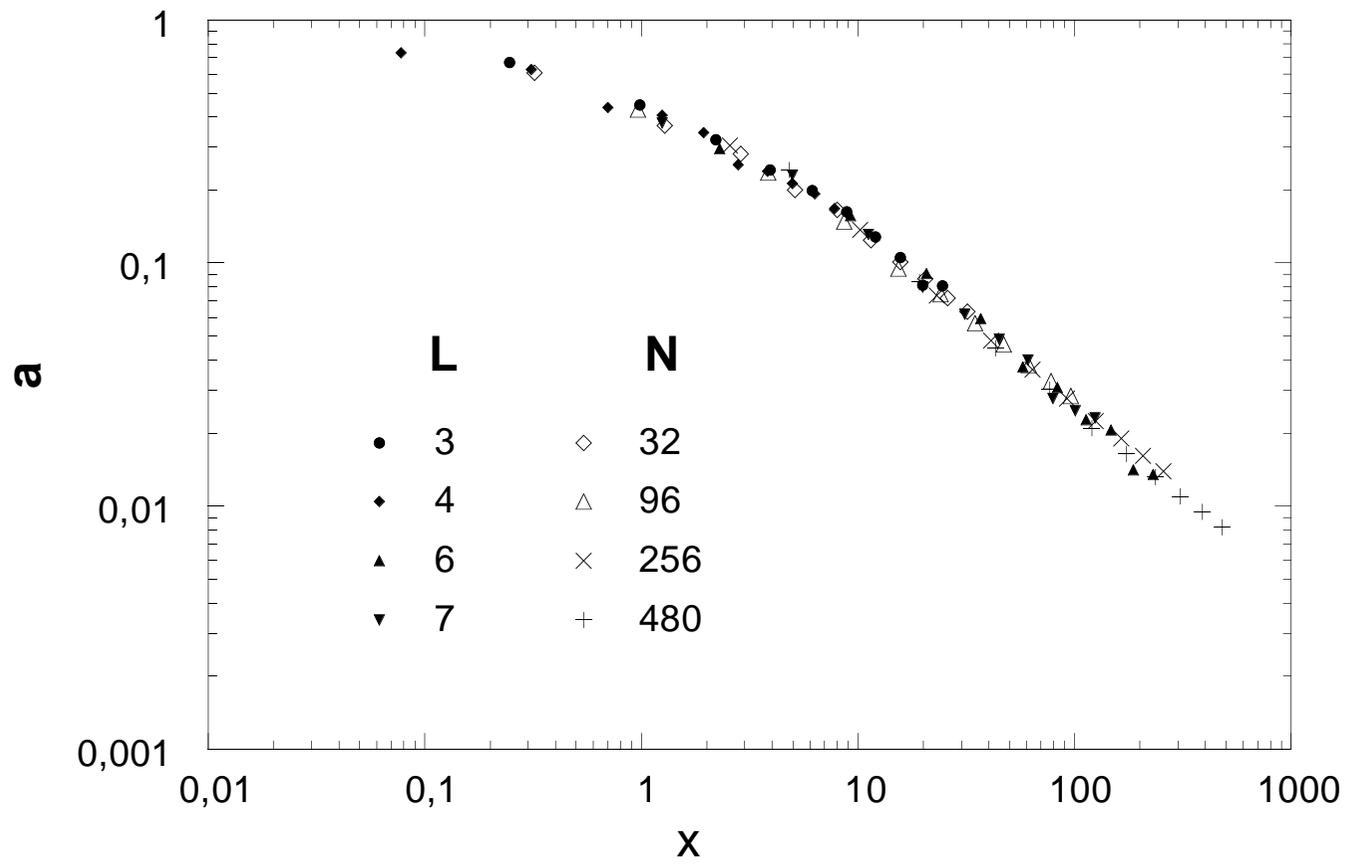
\end{document}